\documentclass[prc,aps,nofootinbib,superscriptaddress,showkeys,showpacs,twocolumn,10pt,floatfix]{revtex4-1}
\usepackage{epsfig}
\usepackage{graphicx}
\usepackage[small]{subfigure}
\usepackage{setspace}
\usepackage[T1]{fontenc}
\usepackage[utf8]{inputenc}
\usepackage[english]{babel}
\usepackage{mathrsfs}
\usepackage{amssymb}
\usepackage{amsmath}
\usepackage{siunitx}
\usepackage{amsopn}
\usepackage{mathtools}
\usepackage{float}
\usepackage{chemformula}
\usepackage{comment} 
\usepackage{tikz}
\usepackage{booktabs}
\usepackage{multirow}
\usepackage{verbatim}
\usepackage{ulem}
\usepackage{bbold}
\usepackage{soul}
\usepackage{upgreek}
\usetikzlibrary{arrows}

\begin{document}

\title{Probing nuclear structure and the equation of state \\ through pre-equilibrium dipole emission in charge-asymmetric reactions}

\author{L. Shvedov} 
\email[]{shvedov@lns.infn.it}
\affiliation{Laboratori Nazionali del Sud, INFN, I-95123 Catania, Italy}
\author{S. Burrello}
\email[]{burrello@lns.infn.it}
\affiliation{Laboratori Nazionali del Sud, INFN, I-95123 Catania, Italy}
\author{M. Colonna}
\email[]{colonna@lns.infn.it}
\affiliation{Laboratori Nazionali del Sud, INFN, I-95123 Catania, Italy}
\author{H. Zheng} 
\email[]{zhengh@snnu.edu.cn}
\affiliation{School of Physics and Information Technology, Shaanxi Normal University, Xi'an 710119, China}

\begin{abstract}
We investigate the pre-equilibrium dipole response in the charge-asymmetric reaction $^{40}$Ca~+~$^{152}$Sm, of recent
experimental interest, at several beam energies within the range $[5, 11]$ AMeV and different collision centralities. By employing Skyrme-like effective interactions for the nuclear mean field, we probe the role of the different ingredients performing theoretical calculations based on the time-dependent Hartree-Fock approach or a semi-classical transport model that also includes two-body correlations.  A comparative analysis between these approaches allowed us to disentangle the role of deformation effects in the entrance channel from the ones associated with structure details of genuine quantal nature on the dipole emission. Moreover, we also investigate the impact of the occurrence of residual two-body collisions on the reaction dynamics. This study contributes to the understanding of the microscopic processes that determine the complex dynamics of low-energy heavy-ion collisions along the fusion-fission path, which is relevant to super-heavy element synthesis, unraveling  interesting connections with the characteristics of the nuclear effective interaction and the associated equation of state.
\end{abstract}


\maketitle

\section{Introduction}
Low-energy nuclear reactions span a variety of competing processes, ranging from (incomplete) fusion to binary exit channels, such as quasi-fission or deep-inelastic scattering~\cite{LacroixPPNP2004, NakatsukasaRMP2016}. The study of the related reaction mechanisms may provide crucial information about energy sharing and dissipation among the colliding nuclei, which finally determine the extent of equilibration reached during the reaction path and the properties of the reaction products~\cite{barPR2005, WilliamsPRL2018, zhengPRC2018}. These features are generally determined by an intricate balance between single-particle behavior and the potential emergence of collective motion, reflecting the nuanced nature of the self-consistent mean-field and the possible impact of shell effects~\cite{benderRMP2003, WashiyamaPRC2009, burrelloPRC2019}. 

To enrich the scenario, new interesting phenomena emerge in reactions involving nuclei with different neutron-to-proton number (N/Z) ratios. In such cases, the centers of mass  of protons and neutrons do not coincide during the early stage of the collision, and charge equilibration can be reached either through incoherent exchange of nucleons between the reaction partners or through the excitation of a collective
dipole oscillation mainly along the symmetry axis of the highly deformed composite di-nuclear system~\cite{ChomazNPA1993, BaranNPA1996, FlibottePRL1996, SimenelPRL2001}. This motion, referred to as the dynamical dipole (DD) mode~\cite{PierroutsakouEPJA2003, PierroutsakouPRC2005, PierroutsakouPRC2009, WuPRC2010}, is responsible for the emission of  pre-equilibrium dipole radiation during the di-nuclear phase of the reaction, to be distinguished from the  statistical $\gamma$-ray emission expected from the decay of the  giant dipole resonance excited in the compound nucleus at a later stage~\cite{SimenelPRC2007, BraccoPPNP2019, LvPRC2021}.  

The DD mode provides crucial insights into the shape and charge distribution of the composite system~\cite{BaranNPA2001} and may serve as a cooling mechanism for the formation of super-heavy elements during the fusion process~\cite{ItkisNPA2015, OganessianRPP2015}, favoring the compound nucleus survival against fission. As evidenced by recent theoretical and experimental analyses, the DD mode is expected to be sensitive to several ingredients, such as charge and mass asymmetry, collision centrality and collision energy~\cite{PapaPRC2005, MartinPLB2008, CorsiPLB2009, GiazPRC2014, Parascandolo:2016ntb}. 
At optimal beam energies ($E_{\rm{\rm lab}}\sim$ 10 AMeV), pre-equilibrium emission of nucleons and light particles can also occur, contributing to the cooling of the system and potentially decreasing the initial charge asymmetry, particularly through the preferential emission of neutrons in neutron-rich systems~\cite{zhengPLB2017}. 

The restoring force driving DD oscillations is provided by the isovector channel of the nuclear effective interaction~\cite{BaranNPA2001, zhengPLB2017}, which ties the DD mode to the symmetry energy of the nuclear Equation of State (EoS), a subject of extensive current research{, also in connection with astrophysical phenomena~\cite{HuthNAT2022, tsangNAT2024}}. In particular, the DD mechanism, as well as the N/Z ratio of the pre-equilibrium nucleon emission, have been
proposed as probes for the low-density behavior of the symmetry energy~\cite{BaranPRC2009, YePRC2013, burrelloFRO2019}, a regime  of crucial importance for nuclear structure features (such as the neutron skin thickness in neutron-rich nuclei)~\cite{CentellesPRL2009, zhengPRC2016, burrelloPRC2021} and for the description of
low-density clustering in compact stars~\cite{typelPRC2010, burrelloPRC2015, burEPJA2022}.

However, in addition to the well-established role of the nuclear symmetry energy, other components of the effective interaction, such as those related to the momentum-dependent part or surface terms, may also significantly influence the main features of the DD emission. In this work, we aim at providing a comprehensive study of these factors, essential for making reliable predictions about the likelihood of observing a robust dipole oscillation in a composite di-nuclear system. Within such a context, we also investigate the possible impact of structure effects, such as nuclear deformation, on the dipole dynamics. Indeed, nuclear deformation is known to influence significantly various aspects of low-energy reaction dynamics, including the magnitude of the fusion cross section at energies near the Coulomb barrier~\cite{UmarPRC2014, UmarPRC2016, zhengPRC2018}.

Another aspect to bear in mind is that, although reactions at energies slightly above the Coulomb barrier are largely dominated by one-body dissipation mechanisms, two-body correlations beyond the mean-field approximation can be important~\cite{zhengPLB2017, LoeblPRC2012}. These are usually accounted for within semi-classical transport approaches, in terms of nucleon-nucleon scattering representing the effect of the hard core of the nuclear interaction. In this study, we consider the charge-asymmetric reaction $^{40}$Ca + $^{152}$Sm, of recent experimental interest~\cite{Parascandolo:2016ntb, parascandoloPRC2022}, at different beam energies within the range [5, 11]
AMeV. By refining semi-classical transport models based on the Boltzmann-Nordheim-Vlasov (BNV)~\cite{BonaseraPR1994, BaranNPA2001} approach, we revisit the role of the different ingredients of the effective interaction and investigate the impact of ground state deformation and surface effects on the {dynamical} dipole emission. In addition, we also consider fully quantum mechanical calculations using the time-dependent Hartree-Fock (TDHF) approach~\cite{LacroixPPNP2004, SchunckRPP2016, SimenelPPNP2018, sekizawaFRO2019}, to disentangle global deformation effects, characterizing the ground-state configuration of $^{152}$Sm, from 
genuine quantal (shell) effects
and better assess their influence on the reaction dynamics. More in general, by employing two microscopic approaches—BNV and TDHF models—and using various energy density functionals to describe the nuclear mean-field, under different reaction configurations, we aim at improving our understanding of the microscopic processes occurring in low-energy Heavy Ion Collisions (HICs), also seeking for connections with the properties of the nuclear effective interaction and the associated EoS~\cite{ColonnaPPNP2020, sorensenPPNP2024}. This paper is organized as follows: in Section~\ref{sec:theo}, the theoretical framework is briefly introduced. The results of the DD emission in the reaction $^{40}$Ca + $^{152}$Sm are shown and discussed in Section~\ref{sec:results}. The conclusions are drawn in section~\ref{sec:conclusions}.

\section{Theoretical framework}
\label{sec:theo}
From a theoretical point of view, understanding the dynamics of low-energy HICs involves solving a complex quantum many-body problem, which is still currently unaffordable in the general case. As such, for several practical purposes,  an important simplification is to resort to the mean-field (MF) approximation, or suitable extensions, thus reducing to a one-body problem for the reaction dynamics. Such an approximation implies the use of so-called effective interactions, which govern to a large extent the collision dynamics, beside the delicate interplay with many-body correlations beyond the MF picture.

\subsection{Mean-field models: TDHF and its semi-classical limit}
Within the independent particle picture, a widely employed approximation corresponds to consider a Slater determinant for the many-body state, leading to the TDHF theory, widely employed in nuclear physics to describe various aspects of nuclear dynamics~\cite{SchunckRPP2016, zhengPRC2018, SimenelPPNP2018, sekizawaFRO2019}. In the TDHF theory, the evolution of the one-body density matrix $\hat{\rho}$ is determined by 
\begin{equation}
i\hbar \partial_{t} \hat{\rho}=\left[\hat{h}[{\rho}],\hat{\rho} \right], 
\label{EQ:TDHF}
\end{equation}
where $\hat{h}[{\rho}]=\dfrac{\textbf{p}^{2}}{2m} +U[{\rho}]$ is the single-particle Hamiltonian and $U$ is the self-consistent MF potential.

However, for the description of heavy-ion reaction dynamics, especially when Fermi or intermediate energies are concerned, the semi-classical limit of the TDHF equations is often considered. Within such a scheme, one describes the time evolution of the nucleon phase-space distribution function, which is nothing but the semi-classical analog of the Wigner transform of the one-body density matrix~\cite{BertschPPNP1980}. Expliciting the two nucleonic species, one has to solve two coupled Vlasov equations~\cite{barPR2005}:
\begin{equation}
\partial_{t} f_q + \nabla_\mathbf{r} f_{\rm q} \cdot \nabla_\mathbf{p} \epsilon_{q}- \nabla_\mathbf{p} f_{q} \cdot \nabla_\mathbf{r} \epsilon_{q} = 0,
\label{vlasov}
\end{equation}
for the neutron and proton distribution functions $f_q({\bf r},{\bf p},t)$, with $q=n,p$, respectively. In the above equations, $\epsilon_{q}$ represents the neutron or proton single-particle energy, which contains the MF potential $U_{q}$. While the Vlasov approach is unable to account for shell effects, it is well suited to describe robust quantum modes, of zero-sound type, in both nuclear matter and finite nuclei~\cite{barPR2005,Baran:2011xg}. Additionally, the model provides an accurate depiction of the reaction dynamics at energies around the Coulomb barrier, accounting for the transition from fusion to quasi-fission and deep-inelastic processes.

However, at increasing beam energies, residual two-body correlations beyond the MF picture, are expected to play a relevant role. To take their effect into account, a Pauli-blocked collision term $I_{\rm coll}[f_n, f_p]$ is introduced in the r.h.s. of Eq.~\eqref{vlasov}, turning the Vlasov approach into the BNV model. 
To treat the collision integral, in our calculations we employ the (energy, angular, and isospin dependent) free nucleon-nucleon (n-n) cross section.

\subsection{Skyrme-like energy density functional}
The effective interaction, or equivalently the {energy density functional (EDF)} $\mathscr{E}$, is then the common ingredient between the semi-classical and quantum MF models. Considering a standard Skyrme interaction, the EDF is expressed in terms of the local isoscalar, $\rho=\rho_n+\rho_p$, and isovector, $\rho_{3}=\rho_n-\rho_p$,  densities and  kinetic energy densities ($\tau=\tau_{n}+\tau_{p}, \tau_{3}=\tau_{n}-\tau_{p}$) as~\cite{burrelloPRC2019}:
\begin{eqnarray}
\mathscr{E}[\rho, \rho_{3}, \tau, \tau_{3}]&\equiv& \mathscr{E}_{\rm kin}(\tau)+ \mathscr{E}_{\rm pot}(\rho, \rho_3, \tau, \tau_3)\nonumber\\
&=&\frac{\hbar^2}{2 m}\tau + C_0\rho^2 + D_0\rho_{3}^2 + C_3\rho^{\sigma + 2} + D_3\rho^{\sigma}\rho_{3}^2  \nonumber\\
&& + C_{\rm eff}\rho\tau + D_{\rm eff}\rho_{3}\tau_{3} \nonumber\\
&& + C_{\rm surf}(\bigtriangledown\rho)^2 + D_{\rm surf}(\bigtriangledown\rho_3)^2,
\label{eq:rhoE}
\end{eqnarray}
where $m$ is the nucleon mass and the coefficients $C_{..}$, $D_{..}$ and $\sigma$ are combinations of {standard} Skyrme parameters.  In particular, the terms with coefficients $C_{\rm eff}$ ($C_{\rm surf}$) and $D_{\rm eff}$ ($D_{\rm surf}$) are the momentum dependent (surface) contributions to the nuclear effective interaction. 
The spin-orbit and the spin densities-dependent terms (not shown in Eq.~\eqref{eq:rhoE} and omitted in semi-classical calculations) are also considered in the quantal approaches. Moreover, the Coulomb interaction is included in all models. 

In our study, we adopt the recently introduced SAMi-J and SAMi-m Skyrme effective interactions~\cite{Roca-Maza:2012dhp, Roca-Maza:2012uor}, which were devised to provide a good description of isospin and spin-isospin resonances, together with well known empirical data such as masses, radii and important nuclear excitations. 
\begin{figure}[tbp!]
\begin{center}
\begin{tabular}{c}
\includegraphics*[width=.48\textwidth]{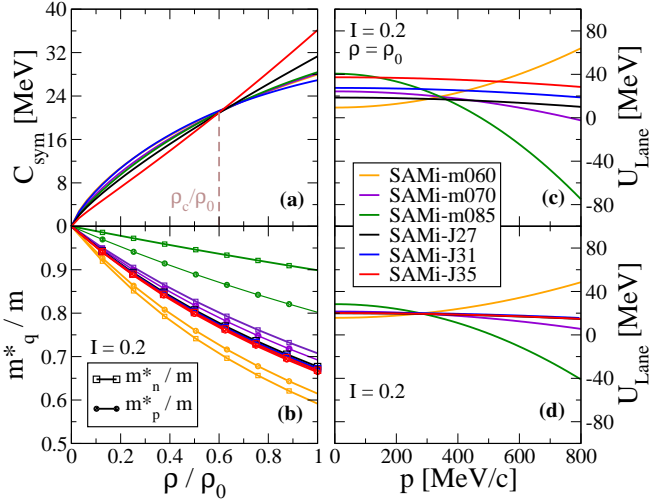}
\end{tabular}
\end{center}
\caption{The symmetry energy coefficient $C_{\rm sym}$ (panel (a)) and the neutron $m_{n}^{\ast}/m$ and proton $m_{p}^{\ast}/m$ reduced effective mass (panel (b)) as function of the reduced isoscalar density $\rho/\rho_{0}$, as obtained by employing the different SAMi EDFs considered in our study. For the same EDFs, the Lane potential as a function of the momentum ${\rm p}$ is shown, as obtained for $I = 0.2$, at saturation density $\rho_{0}$ (panel (c)) and at the crossing density $\rho_{c} = 0.6\rho_{0}$ (panel (d)), respectively.}
\label{fig:sami}
\end{figure}

One should note that, at the equilibrium limit, the EDF, Eq.~\eqref{eq:rhoE}, yields the nuclear EoS, linking the energy per nucleon to neutron and proton densities.  It is worthwhile to introduce the symmetry energy per nucleon, $\dfrac{E_{\rm sym}}{A} = C_{\rm sym}(\rho) I^2$, where $I = \dfrac{\rho_3}{\rho}$ is the asymmetry parameter and the coefficient $C_{\rm sym}$ can be written as a function of the Skyrme coefficients:
\begin{equation}
C_{\rm sym} = \frac{\epsilon_{\rm F}}{3} + D_0\rho + D_3\rho^{\sigma+1} +\frac{2m}{\hbar^2}\left(\frac{C_{\rm eff}}{3} + D_{\rm eff}\right)\epsilon_{\rm F}\rho,
\end{equation}
with $\epsilon_{\rm F}$ denoting the Fermi energy at density $\rho$. Figure~\ref{fig:sami}, panel (a), shows the symmetry energy coefficient $C_{\rm sym}$ as a function of the (reduced) isoscalar density $\rho/\rho_{0}$, being $\rho_{0} = 0.16$ fm$^{-3}$ the saturation density, for the different SAMi EDFs employed in our study. It is important to note that the SAMi-J interactions imply a correlation between the symmetry energy $J = C_{\rm sym}(\rho_{0})$ and its slope $L = 3 \rho_{0} \left. \dfrac{dC_{\rm sym}}{d\rho}\right|_{\rho = \rho_{0}}$ at the saturation density, so that all interactions lead to the same value of the symmetry energy below normal density (at $\rho \approx \rho_c = 0.6\rho_{0}$), {providing} a good reproduction of the ground state (g.s.) properties of nuclei. The SAMi-m effective interactions
exhibit an intermediate trend between SAMi-J27 and SAMi-J31, see Fig.~\ref{fig:sami}, panel (a). 

In the following, we will also make use of the definition of isoscalar $m_{S}^{\ast} = (m_{n}^{\ast} + m_{p}^{\ast})/2$ and isovector effective mass $m_{V}^{\ast} = m_{n}^{\ast} - m_{p}^{\ast}$, being
\begin{equation}
m_{q}^{\ast} = \dfrac{\hbar^2}{2} \left( \dfrac{ \partial \mathscr{E}}{\partial \tau_{q}}\right)^{-1}, \qquad q = n, p    
\end{equation}
the nucleon effective mass, related to the momentum-dependent part of the nuclear effective interaction. The isovector effective mass is actually linked to the derivative, with respect to the momentum ${\rm p}$, of the Lane potential, which has a more intuitive physical meaning. Recalling the expression of the MF potential, $U_{q}= \left. \dfrac{\partial\mathscr{E}_{pot}}{\partial \rho_q} \right|_{\tau, \tau_{3}}$, where $\mathscr{E}_{pot}$ is the potential part of the EDF (Eq.~\eqref{eq:rhoE}), the Lane potential is expressed as~\cite{zhengPRC2018}:    
\begin{eqnarray}
U_{\rm Lane}&=& \frac{U_n-U_p}{2I}. 
\label{lanep}
\end{eqnarray}
Its derivative with respect to the momentum ${\rm p}$ reads:
\begin{equation}
\left. \frac{dU_{\rm Lane}}{d{\rm p}}\right|_{\rho, I}=\frac{\rm p}{2I}
\left(\frac{1}{m_n^*}-\frac{1}{m_p^*}\right)
\label{dudp}
\end{equation}
thus providing a measure of the neutron-proton effective mass splitting~\cite{zhangPLB2015}.

Panel (b) of Fig.~\ref{fig:sami} shows the (reduced) nucleon effective mass  $(m_{q}^{\ast}/m)$, as a function of the (reduced) isoscalar density $\rho/\rho_{0}$, for the asymmetry parameter $I = 0.2$. From panel (b), it clearly emerges that the SAMi-J interactions, which keep unaltered the isoscalar channel, predict nearly identical values for the neutron and proton effective masses (the corresponding curves indeed almost overlap). As a result, the isovector effective mass turns out to be practically zero, while the isoscalar effective mass at saturation density is equal to $m_{S}^{\ast}(\rho_{0})/m = 0.67$. Differently, the SAMi-m effective interactions are characterized (and labeled) according to their isoscalar effective mass value at saturation density, which increases progressively from SAMi-m060 to SAMi-m085. At the same time, the isovector effective mass shifts from negative to significantly positive values. To better illustrate the differences in the isovector mass, panels (c) and (d) of Fig.~\ref{fig:sami} display the Lane potential as a function of the momentum ${\rm p}$, as obtained with the SAMi EDFs employed in our study, for the illustrative cases of $\rho = \rho_{0}$ and  $\rho = \rho_{c}$, respectively, and an asymmetry parameter $I = 0.2$. One observes that, while the SAMi-J EDFs are associated with a relatively flat momentum dependence of the Lane potential, the trend changes from monotonically increasing to strongly decreasing as one moves from SAMi-m060 to SAMi-m085. This behavior also implies that the Lane potential at low momentum increases progressively from SAMi-m060 to SAMi-m085. 

As we will see in Section~\ref{sec:results}, each ingredient of the nuclear effective interaction, including the surface contributions, plays a specific role in shaping the main features of the nuclear response taking place along the reaction dynamics.

\subsection{Numerical details}
The two colliding nuclei are
initialized separately and positioned according to the desired impact parameter $b$, at an  initial distance between their centers of mass (c.m.) of $d = 24.0$ fm. The reaction dynamics is then followed
until a final time of $t_{\rm max} = 630$ fm/c. 

In the quantal calculations, we use the EV8 Hartree-Fock (HF) code~\cite{BoncheCPC2005} to initialize the two nuclei, as a preliminary step before running the TDHF code~\cite{LacroixPPNP2004, KimJPG1997} to follow the time evolution of the reaction system. For that purpose, we adopt a three-dimensional ($96\times 40\times 16$) lattice mesh with a mesh step of $\Delta x = 0.8$ fm and a time step $\Delta t = 0.3 $ fm/c~\cite{ScampsPRC2013}.

In the semi-classical case, the g.s. configuration is given by the stationary solution of Eq.~\eqref{vlasov}, and the integration of the transport equations is based on the test-particle (t.p.) (or pseudo-particle) technique suggested by Wong~\cite{Wong:1982zzb}. We employ $600$ t.p. per nucleon, ensuring a good spanning of the phase space. We note here that the t.p. method is able to reproduce accurately the EoS of nuclear matter and provide reliable results regarding g.s. properties of finite nuclei~\cite{Idier:1993pk, Schuck:1988uoq, BaranPRC2013}. In practice, the  numerical procedure to produce (spherical) g.s. consists in distributing neutrons and protons inside spheres with radii $r_{n}$ and $r_{p}$, respectively. Accordingly, particle momenta are initialized inside Fermi spheres associated with the local neutron or proton densities. Then $r_{n}$ and $r_{p}$ are tuned in order to minimize the total energy, corresponding to the adopted effective interaction. The t.p. method often implies the use of finite width wave packets. For example, the triangular functions are adopted in the present study. Then, some surface effects are automatically included in the initialization procedure and in the dynamics, even without considering explicit surface terms, as those contained in the Skyrme effective interaction. Indeed, for the nuclei selected in our analyses, a good reproduction of the experimental values of the proton root-mean-square radius $\langle r_{p}^{2} \rangle^{1/2}$ (Eq.~\eqref{eq:rsquare}) and the binding energy per nucleon $E_{\rm bind}/A$ is obtained when taking  $C_{\rm surf} = D_{\rm surf} = 0$ (see Eq.~\eqref{eq:rhoE}) in the considered parametrizations (see Table~\ref{tab:bind_radius}).  Therefore, this choice will generally be adopted throughout the study within the semi-classical approaches, unless explicitly stated otherwise. The treatment of surface effects requires indeed special attention, and a dedicated analysis will be presented in Section~\ref{sec:results}.

\subsection{Ground state deformation and reaction configurations}
Let us concentrate hereafter on the charge-asymmetric reaction $^{40}$Ca + $^{152}$Sm. Regarding the g.s. shape of the systems under investigation, while the doubly closed-shell $^{40}$Ca is obviously spherical, nuclear structure studies using various experimental techniques~\cite{Bertozzi:1972zz, Shamu:1976hcn} have shown that $^{152}$Sm is prolate, with deformation parameter values $\beta_{2}$ varying within the range of $[0.220, 0.287]$. It is well known that the orientation of deformed nuclei can significantly affect the exit channel and outcomes of low-energy nuclear reactions near the Coulomb barrier~\cite{UmarPRC2016, zhengPRC2018}. Therefore, we could expect an impact also on the pre-equilibrium dipole radiation. 

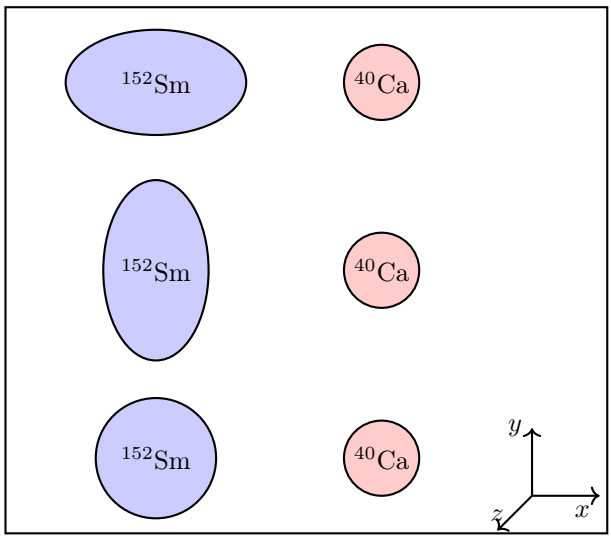
\begin{figure}
\begin{tikzpicture}
\draw[thick] (-9, -0.5) rectangle (-1, 6.5);
    \fill[red!20] (-4, 5.5) circle (.5cm);
    \node at (-4, 5.5) {${}^{40}\text{Ca}$};
    \draw[thick] (-4,5.5) circle (.5cm);
    \fill[blue!20] (-7, 5.5) ellipse (1.2cm and .7cm);
    \node at (-7, 5.5) {${}^{152}\text{Sm}$};
    \draw[thick] (-7,5.5) ellipse (1.2cm and 0.7cm);
    \fill[red!20] (-4, 3) circle (.5cm);
    \node at (-4, 3) {${}^{40}\text{Ca}$};
    \draw[thick] (-4,3) circle (.5cm);
    \fill[blue!20] (-7, 3) ellipse (.7cm and 1.2cm);
    \node at (-7, 3) {${}^{152}\text{Sm}$};
    \draw[thick] (-7, 3) ellipse (.7cm and 1.2cm);
    \fill[red!20] (-4, 0.5) circle (.5cm);
    \node at (-4, 0.5) {${}^{40}\text{Ca}$};
    \draw[thick] (-4,0.5) circle (.5cm);
    \fill[blue!20] (-7, 0.5) circle (.8cm);
    \node at (-7, 0.5) {${}^{152}\text{Sm}$};
    \draw[thick] (-7,0.5) circle (.8cm);
    \begin{scope}[shift={(-2,0)}, scale=0.75] 
        \draw[thick,->] (0,0,0) -- (1.2,0,0) node[anchor=north east]{$x$}; 
        \draw[thick,->] (0,0,0) -- (0,1.2,0) node[anchor= east]{$y$}; 
        \draw[thick,->] (0,0,0) -- (0,0,1.6) node[anchor=south] {$z$}; 
     \end{scope}
\end{tikzpicture}
    \caption{Schematic representation of the three different reaction configurations considered in our study, according to the three possible projectile-target orientations, which correspond to $\langle y^2\rangle=\langle z^2\rangle<\langle x^2\rangle$ (top),  $\langle x^2\rangle=\langle z^2\rangle<\langle y^2\rangle$ (center) and $\langle x^2\rangle=\langle y^2\rangle<\langle z^2\rangle$ (bottom). \label{fig:configurations}}
\end{figure}

Since the deformation of $^{152}$Sm is naturally addressed within the HF approach, the latter is then adopted here as a benchmark to refine the semi-classical models, in the direction of including the deformation effects in the g.s. configuration. The proposed technique consists in rescaling the t.p. coordinates, in order to reproduce the shape of a spheroid, whose surface (considering $z$ as the rotational symmetry axis) is described by the equation: 
\begin{equation}
\frac{x^2+y^2}{b^2} + \frac{z^2}{a^2}=1, \label{spheroideq}
\end{equation}  
where $a=r_{0}(1+s)$ and $b=r_{0}(1-s)$, being $r_{0}=1.2A^{1/3}$, with $A$ denoting the mass number of the nucleus considered and $s$ the scaling parameter. Three reaction configurations are then considered, which correspond to three possible projectile-target orientations, i.e., $\langle x^2\rangle=\langle y^2\rangle<\langle z^2\rangle$,   $\langle x^2\rangle=\langle z^2\rangle<\langle y^2\rangle$ and  $\langle y^2\rangle=\langle z^2\rangle<\langle x^2\rangle$ (see Fig.~\ref{fig:configurations}), where
\begin{equation}
\langle r_{i}^2 \rangle = \dfrac{\int d\textbf{r} r_{i}^2 \rho({\bf r})}{\int d\textbf{r} \rho({\bf r})},
\label{eq:rsquare}
\end{equation}
and $r_{i}$ stands for $x, y, z$ with $i=1, 2, 3$, respectively. It is worthwhile to notice that, in the BNV code, $\vec{x}$ and $\vec{y}$ correspond to the fixed directions of the beam and impact parameter axes, respectively. However, in the adopted TDHF code~\cite{KimJPG1997}, both beam and impact parameter axes are not fixed, being rotated by the angle $\alpha = \arcsin{b/d}$, to optimize the dimension of the calculation box.

The quadrupole moment {$Q_{20}$} of the deformed nucleus can be easily derived (see Appendix~\ref{app:quadrupole}) by assuming a uniform density inside the nuclear surface defined by Eq.~\eqref{spheroideq}:
\begin{equation}
Q_{20} = \int d^3r (2z^2 - x^2 - y^2)\rho({\bf r}) = \frac{8}{5}Ar_{0}^2 s.\label{qmeq1}
\end{equation}
On the other hand, the quadrupole moment can also be expressed in terms of the deformation parameters $\beta_{2}$ and $\beta_{4}$~\cite{Cwiok:1996kn} as
\begin{equation}
Q_{20}=\frac{3}{\sqrt{5\pi}}Ar_{0}^2 \left(\beta _2+\frac 2 7\sqrt{\frac 5{\pi }}\beta _2^2+\frac{12}{7\sqrt{\pi }} \beta _{2}\beta_{4} +\frac{20}{77}\sqrt{\frac 5{\pi }}\beta_{4}^{2} \right).
\label{qmeq2}
\end{equation}
Since the deformation parameters $\beta_{2}$ and $\beta_{4}$ are typically very small, a suitable approximation would be to retain only first order terms in Eq.~\eqref{qmeq2}. In this way, comparing Eq.~\eqref{qmeq1} and Eq.~\eqref{qmeq2}, the scaling parameter $s$ is easily connected with the deformation parameter $\beta_{2}$, as deduced in HF calculations (or taken from experimental data). It is worth noticing that, in case of axially symmetric shapes and neglecting additional terms depending on the deformation parameter $\beta_{6}$ as in Ref.~\cite{Cwiok:1996kn}, Eq.~\eqref{qmeq2} is equivalent to the expression for the expectation value of the axial quadrupole moment which is adopted in the HF calculation, that is
\begin{equation}
\hat{Q}_{20}\equiv 2r^2 P_2(\cos\theta),
\end{equation}
where %$\theta$ is the usual azimuthal angle and 
$P_{2}$ is the Legendre polynomial and $r(\theta, \phi)$ is defined in Appendix~\ref{app:relations}, whereas the relations between the multipole moments and the deformation parameters are also derived.

\subsection{Bremsstrahlung analysis of the reaction dynamics}
In charge-asymmetric reactions, such as the one considered in this work, the ratios $\left(\dfrac{N}{Z}\right)_{P}$ and $\left(\dfrac{N}{Z}\right)_{T}$ are different for projectile (P) and target (T). This generates a finite isovector dipole moment, which is defined, in coordinate space, as: 
\begin{equation}
D(t)=\dfrac{N_{\rm PT}Z_{\rm PT}}{A_{\rm PT}}(R_p-R_n),
\end{equation}
where $A_{\rm PT}=A_{\rm P}+A_{\rm T}$ is the total mass of the di-nuclear system, $N_{\rm PT}=N_{\rm P}+N_{\rm T}$ ($Z_{\rm PT}=Z_{\rm P}+Z_{\rm T}$) is the neutron (proton) number and $R_p$ and $R_n$ refer to the c.m. of protons and neutrons, respectively. Specifically, the dipole moment at the touching point 
\begin{equation}
D_{0} = \dfrac{N_{\rm T} Z_{\rm P} - Z_{\rm T} N_{\rm P}}{A_{\rm PT}} r_{\rm PT}     
\end{equation}
where $r_{\rm PT} = 1.2 \left( A_{\rm P}^{1/3} + A_{\rm T}^{1/3} \right)$, for the reaction here investigated, is $|D_{\rm 0}| = 30.65$ fm.

The dipole moment can trigger DD oscillations along the rotating reaction symmetry axis, whose time evolution can be easily followed within our dynamical models. As done in previous studies investigating the DD $\gamma$-decay~\cite{BaranPRL2001DD, BaranPRC2009, zhengPLB2017}, we adopt a collective bremsstrahlung analysis. Denoting by $E_\gamma$ the photon energy ($E_\gamma=\hbar \omega$), the 
differential emission probability 
associated with dipole oscillations is given by:
\begin{equation}
\frac{dP}{dE_\gamma} = \frac{2e^2}{3\pi \hbar c^3 E_\gamma} |D^{\prime\prime}(\omega)|^2, \label{gmdecay}
\end{equation}
where $e$ is the electric elementary charge and $D^{\prime\prime}(\omega)$ is the Fourier transform of the dipole acceleration $D^{\prime\prime}(t)$~\cite{BaranNPA2001}. 
Results for the latter quantity will be presented in Section~\ref{sec:results}, as obtained by employing the different microscopic approaches considered in our study, under several reaction configurations and using various
EDFs to describe the nuclear MF.
A first insight into the expected behavior of the emission probability is obtained 
by considering a damped harmonic oscillator. In this case, denoting the oscillation frequency and damping rate by $\omega_0$ and $\uptau_{\rm coll}$, respectively, one can derive the following expression for the Fourier transform of the dipole acceleration~\cite{BaranPRC2009}:
\begin{equation}
 |D^{\prime\prime}(\omega)|^2 = \frac{(\omega_0^2 + 1/\uptau_{\rm coll}^2)^2 D_{0}^2}{(\omega - \omega_0)^2 + 1/\uptau_{\rm coll}^2}. \label{psana}
\end{equation}
From this equation, it is evident that the strength of the DD emission is influenced by the initial dipole amplitude $D_{0}$, while also depending on the oscillation frequency, with higher values of $\omega_{0}$ leading to stronger emission.

\section{Results}
\label{sec:results}
In this section, we will discuss the results, relating to the power spectrum  $|D^{\prime\prime}(\omega)|^2$, which determines the DD $\gamma$-emission probability. Specifically, we concentrate on the charge-asymmetric reaction $^{40}$Ca+$^{152}$Sm at $E_{\rm{\rm lab}} / A = 11$ MeV, which was the object of recent studies~\cite{Parascandolo:2016ntb, parascandoloPRC2022}, where the pre-equilibrium component resulting from DD excitation was identified and measured. In Ref.~\cite{parascandoloPRC2022}, the centroid energy $E_{\rm DD}^{\rm exp}= (11.7\pm 0.3)$ MeV and the width $\Gamma_{\rm DD}^{\rm exp}= (2.0\pm 0.5)$ MeV, describing the experimental DD $\gamma$-spectrum via a Lorentzian distribution, were extracted. Moreover, the experimental energy- and angle-integrated DD $\gamma$-ray multiplicity yield $M_{\gamma, \rm DD}^{\rm exp}$, corrected for the detection efficiency and corresponding to an average impact parameter of $\langle b_{\rm exp} \rangle \sim 5$ fm, were compared with theoretical predictions from the BNV transport model~\cite{BaranNPA2001}, employing a collective bremsstrahlung analysis. The centroid energy predicted by theoretical calculations $(E_{\rm DD}^{\rm th})$—assuming a spherical g.s. for $^{152}$Sm—underestimated $E_{\rm DD}^{\rm exp}$. At the same time, the BNV predictions of the DD $\gamma$-multiplicity $M_{\gamma, \rm DD}^{\rm th}$ overestimated the corresponding experimental value $M_{\gamma, \rm DD}^{\rm exp}$. These discrepancies raise the question whether specific ingredients of the nuclear effective interaction, which were not included in previous analyses, and structure effects, such as nuclear deformation, may play a role. The impact of these ingredients on the DD is investigated in the following.

\subsection{Effect of momentum dependence and symmetry energy}
First, let us discuss the role of some specific ingredients of the nuclear effective interaction, particularly those related to its isovector channel and to the momentum dependence.

In this regard, we build upon previous findings of Ref.~\cite{zhengPLB2017}, which used the BNV model to investigate a different charge-asymmetric reaction, namely  $^{132}$Sn+$^{58}$Ni at $E_{\rm lab}/A=10$ MeV. That study highlighted notable differences in the DD power spectra when using SAMi-J momentum-dependent interactions compared to simplified momentum-independent (MI) interactions ($C_{\rm eff} = D_{\rm eff} = 0, m_{q}^{\ast} = m$), denoted as asysoft, asystiff and  asysuperstiff. These MI interactions offered three distinct types of density dependence, each with a similar symmetry energy value $J \approx 30$ MeV, but differing in the slope parameter $L$ and crossing each other at $\rho = \rho_{0}$~\cite{BaranPRC2013}. 

\begin{figure}[tbp!]
\begin{center}
\begin{tabular}{c}
\includegraphics*[width=8.5cm]{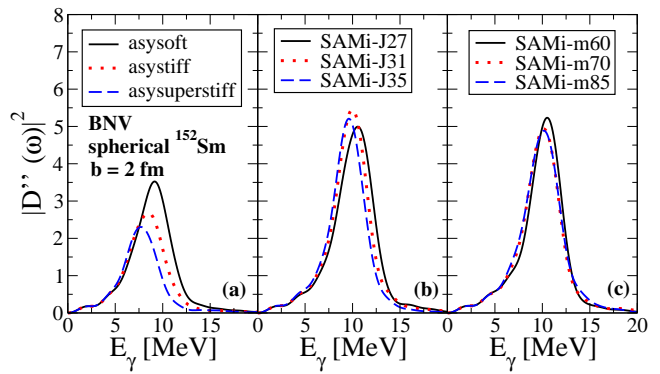}
\end{tabular}
\end{center}
\caption{The power spectra of the dipole acceleration for the reaction $^{40}$Ca+$^{152}$Sm at $E_{\rm{\rm lab}} / A = 11$ MeV employing BNV model with MI (panel (a)) and SAMi-J (panel (b)) or SAMi-m (panel (c)) effective interactions. The impact parameter $b=2$ fm was considered and a spherical configuration was assumed for the g.s. of $^{152}$Sm.}
\label{figure1}
\end{figure}
Figure~\ref{figure1} shows the power spectra for the reaction $^{40}$Ca+$^{152}$Sm at $E_{\rm lab}/A=11$ MeV, obtained from the BNV model at an impact parameter of $b = 2$ fm, for MI (panel (a)) and SAMi-J (panel (b)) or SAMi-m (panel (c)) effective interactions. Consistent with previous findings of Ref.~\cite{zhengPLB2017}, our analysis confirms that, in the MI case, the centroids of the power spectra show a dependence on the specific effective interaction used, whereas in the SAMi case, they remain relatively close to each other. Since the restoring force of isovector dipole oscillations is primarily driven by the symmetry energy, these results confirm that DD effectively probes the density region around the crossing point of the SAMi interactions. The sensitivity to this specific density range reflects the elongated shape of the system during the pre-equilibrium phase, lowering the oscillation frequency with respect to the standard giant dipole resonance (GDR).

Moreover, the power spectra show a larger magnitude, with their centroids shifting to higher energies in the SAMi case compared to the MI one. This result can be linked to the well-known influence of the (isoscalar) effective mass on the properties of collective modes, such as the GDR. In particular, the SAMi interactions result in a higher oscillation frequency and an increase in the Energy Weighted Sum Rule. Thus, the results of Fig.~\ref{figure1} fully reaffirm the conclusions from Ref.~\cite{zhengPLB2017}. A higher strength of the spectra which corresponds to a higher centroid energy is moreover consistent with the behavior of a damped harmonic oscillator, as described by Eq.~\eqref{psana}. Additionally, this finding potentially resolves the issue of $E_{\rm DD}^{\rm th}<E_{\rm DD}^{\rm exp}$ observed with the MI effective interaction in Refs.~\cite{Parascandolo:2016ntb, parascandoloPRC2022}. However, the use of momentum-dependent effective interaction does not have a significant impact on the gamma multiplicity, as we will discuss in the following (see Section~\ref{sec:2body}). 

Interestingly, the expected magnitude ordering of the power spectra based on the isoscalar effective mass values for SAMi-m effective interactions is not observed in panel (c) of Fig.~\ref{figure1}. This apparent inconsistency can be attributed to a delicate interplay between the role of isoscalar and isovector effective masses (or, equivalently, the neutron-proton effective mass splitting). Specifically, the sign of this splitting changes from negative to positive when moving from SAMi-m060 to SAMi-m085 (see panel (b) of Fig.~\ref{fig:sami}). Consequently, in the case of SAMi-m085, the Lane potential exhibits an increasing repulsion at the low momenta experienced in the nuclear reaction (see panels (c) and (d) of Fig.~\ref{fig:sami}). This leads to a shift of the oscillations towards higher energy, which counterbalances the effect of the larger isoscalar effective mass (see panel (c) of Fig.~\ref{figure1}). 

Finally, it is worth noting that similar features about the effects of the momentum dependence and of the symmetry energy were observed also for other impact parameters. The only notable difference is a reduction in the magnitude of the power spectrum in less central collisions. For the sake of illustration, only the SAMi-J31 effective interaction will be then considered hereafter.

\subsection{Effect of ground state deformation}
\begin{figure}[tbp!]
\begin{center}
\begin{tabular}{c}
\includegraphics*[width=8.5cm]{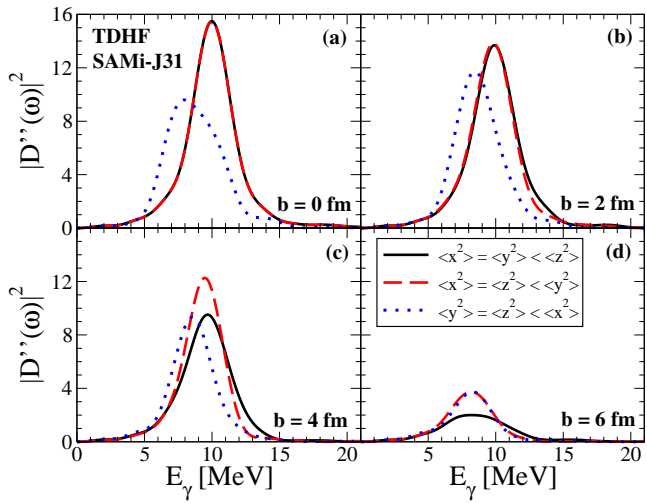}
\end{tabular}
\end{center}
\caption{The power spectra of the dipole acceleration for the reaction $^{40}$Ca+$^{152}$Sm at $E_{\rm{\rm lab}} / A = 11$ MeV with SAMi-J31 from TDHF at different impact parameters and collision configurations.}
\label{fig:TDHF}
\end{figure}

To investigate whether incorporating deformation effects in the g.s. configuration impacts the DD $\gamma$-multiplicity, we turn to the (quantal) TDHF model, where deformation is naturally accounted for.  We consider, for different impact parameters, the collision configurations based on the three possible P-T orientations (see Fig.~\ref{fig:configurations}). The corresponding power spectra are shown in Fig.~\ref{fig:TDHF}.

However, since the three collision configurations have the same probability of occurrence, the experimental power spectra should be compared to the average of the corresponding results. In Fig.~\ref{fig:TDHF_Vlasov}, the averaged power spectra of the DD at different impact parameters, calculated using the TDHF model with the SAMi-J31 interaction, are shown as solid black curves. To highlight the role of the deformation, the black dashed lines represent the case where the energy minimization leading to the g.s. configuration within the variational HF procedure is constrained to maintain the spherical shape of $^{152}$Sm.

By comparing the solid and dashed black lines, one observes that the significant differences in the power spectra corresponding to the various collision configurations shown in Fig.~\ref{fig:TDHF} are strongly mitigated when averaging over the three possible P-T orientations. This compensation effect is particularly evident in central collisions. In this case, the overall effect compared to the spherical case is a shift to higher centroid energies. This result reflects the prolate shape of $^{152}$Sm, which reduces the effective size of the di-nuclear system in two out of the three possible collision configurations, as shown in panels (a) and (b) of Fig.~\ref{fig:TDHF}. In the more peripheral collisions, where the surface of the composite system plays a more prominent relative role, this shift becomes negligible, as observed in panels (c) and (d) of Fig.~\ref{fig:TDHF_Vlasov} when comparing the solid and dashed black lines. Moreover, when the deformation of the g.s. configuration is accounted for, a 
suppression of the power spectrum is observed. However, this suppression emerges only in a narrow window of impact parameters where also other effects likely play a role in this respect. In particular, as we will discuss in the following, we expect the damping rate of DD oscillations to be significantly influenced by two-body n–n collisions~\cite{zhengPLB2017}. Then, to further investigate this effect, we will need to employ the newly devised semi-classical BNV approach that incorporates the g.s. deformation effects. 

\subsection{Comparison between TDHF and Vlasov results}
\label{sec:surface}
Before embarking on the full calculations with BNV, it is interesting to compare the results of its collisionless Vlasov limit with those from TDHF, as the Vlasov model represents the semi-classical counterpart to the fully quantal TDHF approach employed so far. A comparative analysis between semi-classical and quantal approaches, within the same EDF framework, would allow us to assess the extent to which a semi-classical picture can explain the properties of nuclear dynamics~\cite{burrelloPRC2019, burrelloFRO2019}.

\begin{figure}[t]
\begin{center}
\begin{tabular}{c}
\includegraphics*[width=8.5cm]{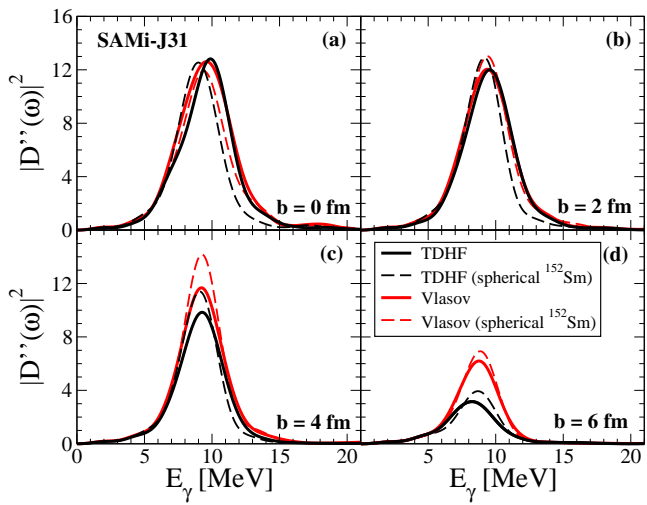}
\end{tabular}
\end{center}
\caption{The averaged power spectra of the dipole acceleration for the reaction $^{40}$Ca+$^{152}$Sm at $E_{\rm{\rm lab}} / A = 11$ MeV with SAMi-J31 from TDHF and Vlasov models at different impact parameters.}
\label{fig:TDHF_Vlasov}
\end{figure}

We performed Vlasov calculations for the three collision configurations at different impact parameters and obtained the averaged power spectra, shown in Fig.~\ref{fig:TDHF_Vlasov} for the SAMi-J31 interaction, as the red full lines. For comparison, in Fig.~\ref{fig:TDHF_Vlasov} we also present, as the red dashed lines, the results from the Vlasov model with spherical g.s. for both colliding nuclei.

First, by comparing the solid and dashed red lines in all panels of Fig.~\ref{fig:TDHF_Vlasov}, one deduces that incorporating deformation effects in the g.s. of $^{152}$Sm within the Vlasov model leads to similar modifications in the power spectra as those already discussed for the fully quantal TDHF case. This supports the validity of the procedure used to embed deformation effects in semi-classical models.

Additionally, from panels (a) and (b) of Fig.~\ref{fig:TDHF_Vlasov}, a comparison between the solid black and red lines shows that the power spectra from the Vlasov model closely match those from the TDHF model in central collisions ($b = 0$ and $2$ fm). This holds true not only when accounting for the deformation of $^{152}$Sm but also for the spherical case. Therefore, this confirms the reliability of the semi-classical approach in accurately describing the reaction dynamics, particularly in central collisions where the bulk properties of the colliding nuclei play the most prominent role. In such cases, the bulk of the reactants drives the fusion path, making the details of the energy level structure less critical. As a result, the TDHF and Vlasov models (both with deformed and spherical g.s.) yield nearly identical outcomes.

However, at larger impact parameters, as in panels (c) and (d) of Fig.~\ref{fig:TDHF_Vlasov}, the power spectra magnitudes from the Vlasov model exceed those from TDHF. The observed differences could be attributed to the different treatment of surface terms in Vlasov and TDHF, which play a more important role in peripheral reactions ($b = 4$ and $6$ fm).

Unlike deformation effects, which only influence results within a narrow impact parameter range (around $b = 4$ fm), the differences in the TDHF and Vlasov power spectra turn out to be significant across all peripheral collisions, where experimental sensitivity is usually heightened. This difference may offer a key to address the overestimation of DD $\gamma$-ray emission predicted by the standard BNV model~\cite{parascandoloPRC2022}, and  deserves therefore further investigation.

\begin{figure}[t]
\begin{center}
\begin{tabular}{c}
\includegraphics*[width=8.5cm]{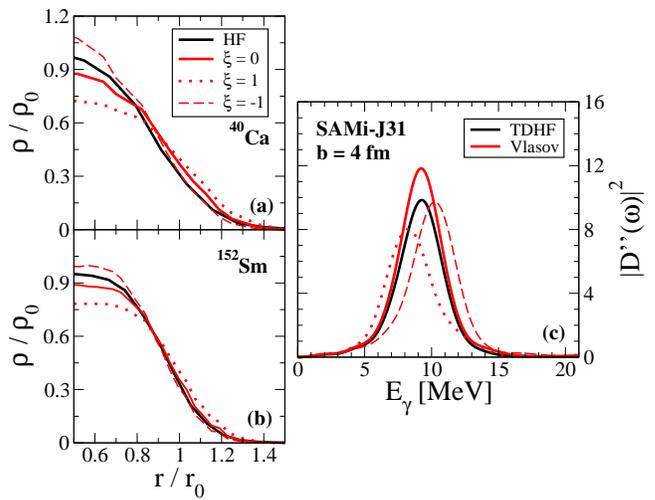}
\end{tabular}
\end{center}
\caption{Left panels: The radial profiles of the reduced isoscalar density, as obtained in the quantal HF approach (black lines) or within the semi-classical procedure (red lines) for $^{40}$Ca (panel (a)) and $^{152}$Sm (panel (b)) with SAMi-J31 effective interaction. Red lines refer to calculations obtained by modulating the surface terms by a scaling coefficient $\xi$. Right panel (panel (c)): The corresponding averaged power spectra of the dipole acceleration for the reaction $^{40}$Ca+$^{152}$Sm at $E_{\rm{\rm lab}} / A = 11$ MeV as obtained from TDHF and Vlasov models at $b = 4$ fm.}
\label{fig:surface}
\end{figure}

\subsubsection{Role of surface terms} As explained in Section~\ref{sec:theo}, in the semi-classical calculations presented so far, we excluded the explicit surface terms from the standard Skyrme-like EDFs (Eq.~\eqref{eq:rhoE}), as global nuclear properties were accurately reproduced with $C_{\rm surf} = D_{\rm surf} = 0$. However, to investigate the role of the surface terms, we now introduce the usual Skyrme contribution and modulate it by a scaling coefficient $\xi$ such that $C_{\rm surf}^{\prime} = \xi C_{\rm surf}$ and $D_{\rm surf}^{\prime} = \xi D_{\rm surf}$. In addition to the case considered so far, corresponding to $\xi = 0$, two other possibilities, namely $\xi = \pm 1$, were explored. In Table~\ref{tab:bind_radius}, we present the binding energy $E_{\rm bind}/A$ and the neutron $\langle r_{n}^{2} \rangle^{1/2}$ and proton $\langle r_{p}^{2} \rangle^{1/2}$ root-mean-square radii, as obtained for $^{40}$Ca and $^{152}$Sm, from both quantal HF and the adopted semi-classical procedure, employing the SAMi-J31 effective interaction. These results are compared with available experimental data for $E_{\rm bind}/A$ and $\langle r_{p}^{2} \rangle^{1/2}$, taken from Refs.~\cite{WangCP2012, AngeliADNDT2013}, respectively. 

\begin{table*}[tbp!]
\centering
\caption{\label{tab:bind_radius} Neutron $\langle r_{n}^{2} \rangle^{1/2}$ and proton $\langle r_{p}^{2} \rangle^{1/2}$ root-mean-square radii (in fm) and binding energy per nucleon $E_{\rm bind}/A$ (in MeV) of the colliding nuclei, as obtained with different models employing SAMi-J31 effective interaction. Experimental data are taken from Refs.~\cite{WangCP2012, AngeliADNDT2013}.}
\begin{tabular}{*{5}{c}} % Adjusted column number
\toprule
 & Nucleus & $\langle r_{n}^{2} \rangle^{1/2}$ [fm] &  $\langle r_{p}^{2} \rangle^{1/2}$ [fm] &  $E_{\rm bind}/A$ [MeV] \\
\midrule
\multirow{2}{*}{\textbf{Experimental data}} 
 & $^{152}$Sm & --- & 5.082 & 8.244 \\
 & $^{40}$Ca & --- & 3.478 & 8.551 \\
\midrule
\multirow{2}{*}{\textbf{HF}} 
 & $^{152}$Sm & 5.200 & 5.047 & 8.148 \\
 & $^{40}$Ca & 3.357 & 3.405 & 8.374 \\
\midrule
\multirow{2}{*}{\textbf{Semi-classical} ($\xi = 0$)} 
 & $^{152}$Sm & 5.273 & 5.072 & 8.515 \\
 & $^{40}$Ca & 3.431 & 3.481 & 8.891 \\
\midrule
\multirow{2}{*}{\textbf{Semi-classical} ($\xi = 1$)} 
 & $^{152}$Sm & 5.478 & 5.266 & 7.130 \\
 & $^{40}$Ca & 3.634 & 3.695 & 6.746 \\
\midrule
\multirow{2}{*}{\textbf{Semi-classical} ($\xi = -1$)} 
 & $^{152}$Sm & 5.080 & 4.900 & 10.123 \\
 & $^{40}$Ca & 3.219 & 3.258 & 11.697 \\
\bottomrule
\end{tabular}
\end{table*}
The accurate reproduction of the experimental data without resorting to explicit surface terms ($\xi = 0$) presented in Table~\ref{tab:bind_radius} is unsurprising, as surface effects were already partially incorporated through the use of t.p. with finite-width wave packets in the initialization and dynamics. Indeed, when surface terms are fully included, as in the standard Skyrme interaction form  ($\xi = 1$), a larger nuclear size is predicted for both systems, along with reduced binding energy due to the positive (repulsive) contribution of the gradient terms. Conversely, in the opposite scenario where surface terms would contribute more strongly to attraction ($\xi = -1$), more compact systems with increased binding energy are predicted.

Panels (a) and (b) of Fig.~\ref{fig:surface} show the radial profile of the reduced isoscalar density $\rho / \rho_{0}$ for $^{40}$Ca (panel (a)) and $^{152}$Sm (panel (b)) using the SAMi-J31 interaction, comparing the quantal HF (black lines) with the semi-classical (red lines)  approach. Without explicit surface terms ($\xi = 0$), the semi-classical density profiles differ from the HF results, particularly in their smoother diffuseness, which enhances surface effects. This is particularly true for the smaller system, $^{40}$Ca, where the relative importance of the surface is heightened in comparison to the bulk. The discrepancy explains why, although the DD centroid energy matches between TDHF and Vlasov calculations, reflecting the nice agreement on the nuclear size predicted by the two approaches,in peripheral collisions the formation of the compound system is more favored in Vlasov than in TDHF, enhancing the DD emission (as shown, e.g. at $b = 4$ fm, in panel (c) of Fig.~\ref{fig:TDHF_Vlasov} or in panel (c) of Fig.~\ref{fig:surface}). Panels (a) and (b) of Fig.~\ref{fig:surface} also clearly demonstrate that surface terms as in the standard Skyrme interaction form  ($\xi = 1$) predict more extended systems with a larger relative contribution from the outer regions. Conversely, the (unrealistic) choice $\xi = -1$ reduces the radii of the nuclei (see Table~\ref{tab:bind_radius}) and makes the semi-classical density profiles sharper than the HF result at the surface, making the systems more compact and reducing the relative importance of the surface contributions. 

In panel (c) of Fig.~\ref{fig:surface}, the corresponding power spectra are plotted as red dotted ($\xi = 1$) or dashed ($\xi = -1$) lines. It is clear that the centroid energies no longer align with the TDHF one due to the different size of the systems, which is larger (smaller) for $\xi = 1$ ($\xi = -1$), respectively. However, the magnitude of the power spectrum at $b = 4$ fm approaches the TDHF result, either for $\xi = 1$ or $\xi = -1$. In the first case, a compensation effect occurs between the enhanced surface effects, driven by the shape of the density profiles and favoring the DD, and the suppression ascribable to both the lower centroid frequency and the larger nucleon emission from the less bound systems. An interesting picture emerges from the case with $\xi = -1$ where more bound systems are predicted. Despite the shift of the centroid energy to higher values, a significant suppression is observed compared to the $\xi = 0$ case, due to the sharper radial transition at the surface, similar to that seen in HF calculations. This finding ultimately highlights the strong  sensitivity of the DD $\gamma$-multiplicity to fine details of the nuclear structure, particularly the nuclear surface, which play a crucial role especially in peripheral collisions. 

However, fine-tuning the surface coefficients within the Vlasov approach to simultaneously reproduce experimental nuclear properties and the DD power spectrum from TDHF is neither simple nor straightforward. Therefore, in the next Section, when comparing with experimental data, we will account for this suppression $a$ $posteriori$, after incorporating the essential effects of two-body correlations through full BNV calculations.

\subsection{Effect of two-body correlations}
\label{sec:2body}
Including residual two-body collisions beyond the MF approximation in the reaction dynamics is crucial to try to reproduce experimental data.

\begin{figure}[tbp!]
\begin{center}
\begin{tabular}{c}
\includegraphics*[width=8.5cm]{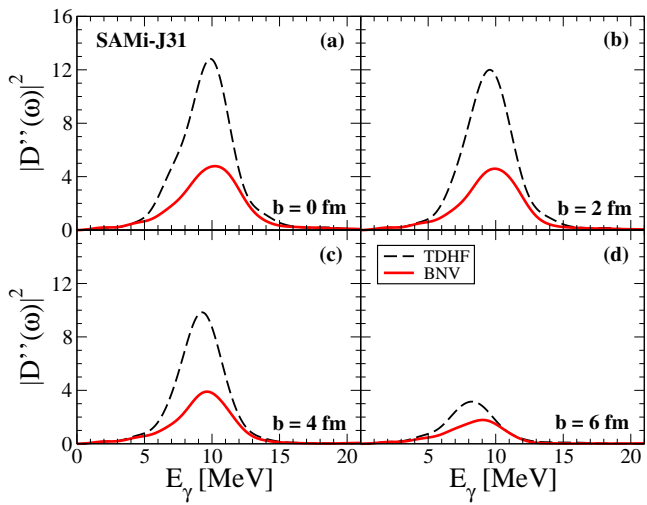}
\end{tabular}
\end{center}
\caption{The averaged power spectra of the dipole acceleration for the reaction $^{40}$Ca+$^{152}$Sm at $E_{\rm{\rm lab}} / A = 11$ MeV with SAMi-J31 from TDHF and BNV models at different impact parameters.}
\label{fig:TDHF_BNV}
\end{figure}

In Fig.~\ref{fig:TDHF_BNV}, we present the averaged power spectra of the DD from the BNV model with deformed g.s. of $^{152}$Sm at different impact parameters. The TDHF results of Fig.~\ref{fig:TDHF_Vlasov}, where no two-body collisions are taken into account, are also shown as a reference. 

It is noteworthy that n-n correlations strongly suppress the power spectra across all impact parameters, leading to a faster thermalization of the di-nuclear system which corresponds to a reduction in the amplitude of the dipole signal. Moreover, unlike surface and deformation effects, the suppression predominantly affects central collisions, as shown in panels (a) and (b) of Fig.~\ref{fig:TDHF_BNV}, while its relative importance decreases with increasing impact parameter, as illustrated in panels (c) and (d). As a result, the dependence of the magnitude of the DD power spectrum on centrality is less pronounced in BNV calculations compared to TDHF results.

On the other hand, the centroids of the power spectra are only marginally influenced by two-body collisions, reaffirming that the DD oscillation frequency is mainly governed by the components of the effective interaction. Notably, a slight shift to higher values is seen in BNV compared to TDHF, particularly in more peripheral collisions, indicating that the system retains its elongated shape for a shorter time when two-body dissipation is present. 

\subsubsection{Comparison with experimental data}
In Table~\ref{tab:multiplicity_exp}, the $\gamma$-multiplicities of the DD emission $M_{\gamma, \rm DD}^{\rm th}$, obtained from the BNV model for the impact parameters $b = 4$ fm and $b = 6$ fm, are compared with the experimental value reported in Ref.~\cite{parascandoloPRC2022}, which corresponds to a mean impact parameter of $\langle b \rangle_{\rm exp} \simeq 5$ fm. For illustrative purposes, the TDHF predictions are also included.

It is evident that, as also discussed in Ref.~\cite{parascandoloPRC2022}, the BNV predictions overestimate the experimental data. The deformation effects introduced here are thus not able to mitigate this problem. 

However, our analysis shows that the structure details of the nuclear surface play a significant role in shaping the DD power spectrum. A possible gateway to effectively incorporate these (quantal) structure effects is to apply to the BNV results a correction factor $\mathtt{x} (b) =\frac{M_{\gamma, \rm DD}^{\rm TDHF} (b) }{M_{\gamma, \rm DD}^{\rm Vlasov}(b)}$, as obtained by comparing, for each impact parameter, the results of the DD $\gamma$-multiplicity in TDHF ($M_{\gamma, \rm DD}^{\rm TDHF}$) and Vlasov ($M_{\gamma, \rm DD}^{\rm Vlasov}$) calculations. As a result, in Table~\ref{tab:multiplicity_exp}, we also present the $\gamma$-multiplicity of the DD emission $M_{\gamma, \rm DD}$ obtained from the BNV model, properly rescaled by the factor $\mathtt{x}$, and denoted as $\mathtt{x}$BNV.

It can be concluded that with this correction, the theoretical predictions of $\mathtt{x}$BNV show much better agreement with the experimental data, significantly reducing the overestimation of $M_{\gamma, \rm DD}$. However, the remaining discrepancy still requires further investigation.

\begin{table}[t]
\centering
\caption{\label{tab:multiplicity_exp} The $\gamma$-multiplicity of the DD emission $M_{\gamma, \rm DD}$ as obtained from TDHF and BNV models, or rescaling the latter by the factor $\mathtt{x}$ ($\mathtt{x}$BNV), for the impact parameters $b = 4$ fm and $b = 6$ fm. The experimental value from Ref.~\cite{parascandoloPRC2022}, corresponding to a mean impact parameter of $\langle b \rangle_{\rm exp} \simeq 5$ fm, is also reported.}
\begin{tabular}{*{5}{c}}
\toprule
 & & \textbf{TDHF} & \textbf{BNV} & \textbf{$\mathtt{x}$BNV} \\
\midrule
\multirow{3}{*}{\textbf{$M_{\gamma, \rm DD}$} [$10^{-3}$]} 
& $b = 4$ fm & 7.59 & 3.63 & 3.02 \\ 
& $b = 6$ fm & 3.15 & 1.99 & 1.25  \\
\cmidrule(lr){2-5}
& $\langle b\rangle_{\rm exp} \simeq 5$ fm & \multicolumn{3}{c}{ 0.82 $\pm$ 0.24} (Ref.~\cite{parascandoloPRC2022}) \\
\bottomrule
\end{tabular}
\end{table}

\subsection{Dependence on the collision energy}
Lastly, an intriguing point to explore, of potential experimental relevance, is the dependence of the results discussed above on the beam energy. 

\begin{figure}[t]
\begin{center}
\begin{tabular}{c}
\includegraphics[width=8.5cm]{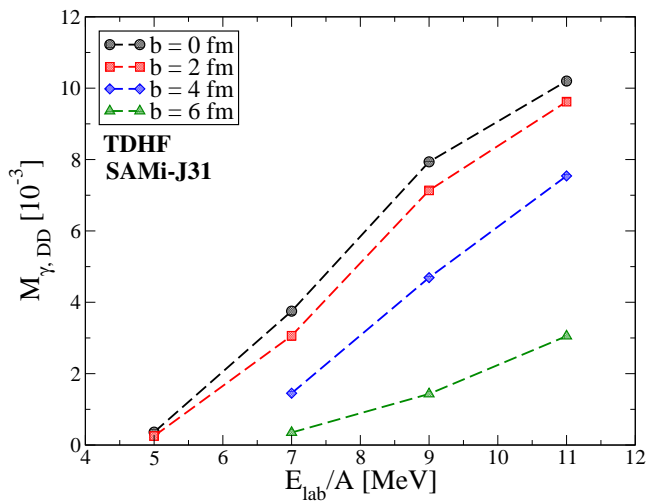}
\end{tabular}
\end{center}
\caption{The $\gamma$-multiplicity of the DD emission $M_{\gamma, \rm DD}$ as a function of the beam energy per nucleon in the laboratory frame $E_{\rm lab}/A$, as obtained by employing the SAMi-J31 effective interaction, with the TDHF model, for different values of the impact parameter $b$. \label{fig:multiplicity_TDHF}}
\end{figure}

For this purpose, Fig.~\ref{fig:multiplicity_TDHF} shows the evolution of the $\gamma$-multiplicity of DD emission, as calculated with the SAMi-J31 effective interaction, as a function of the beam energy per nucleon in the laboratory frame $E_{\rm lab}/A$, for different fixed values of the impact parameter. Although aware of the overestimation of DD $\gamma$-multiplicity, in Fig.~\ref{fig:multiplicity_TDHF} we present only the results from the TDHF model, to provide an upper limit on the minimum beam energy at which collective pre-equilibrium dipole motion might occur.

As a general observation, it is argued that the likelihood of observing a robust dipole oscillation along the fusion path of the composite di-nuclear system increases when moving toward more central collisions, regardless of the beam energy. As a result, quite large values of $M_{\gamma, \rm DD}$ are predicted at the largest beam energy value considered especially for $b = 0$ and $b = 2$ fm. Nevertheless, in (mid-)central collisions, the DD $\gamma$-emission multiplicity strongly decreases while reducing the beam energy. This is likely due to the increased relative role of polarization effects from the Coulomb repulsive interaction between the colliding systems at low beam energies, which hampers the excitation of a collective dipole motion, thereby damping the $\gamma$-ray bremsstrahlung emission. Conversely, for the largest impact parameter considered ($b = 6$ fm), the dependence on the beam energy is milder, because the fusion path (along which the dipole oscillations can take place) is quenched, even at the highest beam energies.  
These results indicate that it is worth isolating central collisions to observe a quite robust DD emission (see, for instance, the analysis of fusion-evaporation events in Ref.~\cite{parascandoloPRC2022}), especially in experiments performed at higher beam energies ($E_{\rm lab}/A \approx 10$ MeV).

\begin{figure}[t]
\begin{center}
\begin{tabular}{c}
\includegraphics[width=8.5cm]{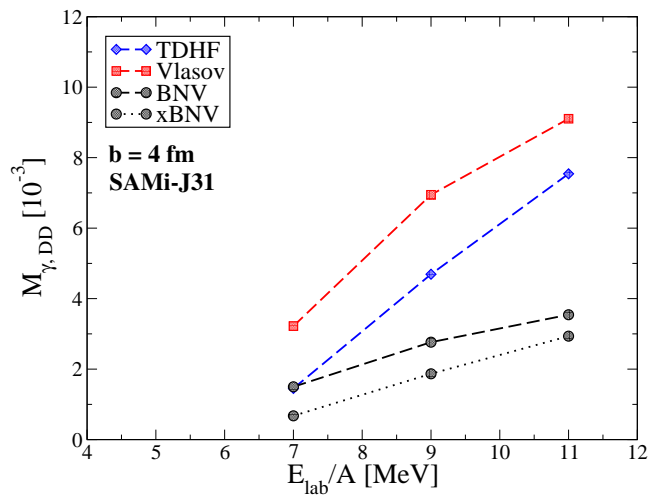}
\end{tabular}
\end{center}
\caption{The $\gamma$-multiplicity of the DD emission $M_{\gamma, \rm DD}$ as a function of the beam energy per nucleon in the laboratory frame $E_{\rm lab}/A$, as obtained by employing the SAMi-J31 effective interaction, with the different models used in our work, for $b = 4$ fm.}
\label{fig:multiplicity_b4}
\end{figure}

However, one should consider that the DD emission is quenched by the effect of n-n correlations. This is illustrated in Fig.~\ref{fig:multiplicity_b4}, which shows the evolution of the $\gamma$-multiplicity of DD emission, calculated at an intermediate impact parameter ($b = 4$ fm) with the SAMi-J31 effective interaction, as a function of beam energy per nucleon, $E_{\rm lab}/A$, across the different theoretical models employed in our study. The steep increase with energy predicted in the collisionsless (Vlasov and TDHF) models is mitigated when the effect of residual two-body collisions is included, resulting in a flatter behavior of $M_{\gamma, \rm DD}$ over the $E_{\rm lab}/A$ range considered.

These findings can guide the experimental choice of the most suitable conditions to detect the DD emission, also in view of the beam energies at which exotic beams are more readily available. 

\section{Conclusions}
\label{sec:conclusions}
In summary, this work explores the pre-equilibrium dipole response in the charge-asymmetric reaction $^{40}$Ca + $^{152}$Sm over a beam energy range of $[5, 11]$ AMeV and different collision centralities. Our analysis provides a detailed examination of the key factors influencing the likelihood of dipole oscillations in the composite di-nuclear system which forms along the fusion path in low-energy HICs, linking this mechanism to the features of the nuclear effective interaction and the associated EoS.

Using Skyrme-like effective interactions for the nuclear MF, we conduct theoretical calculations based on either the quantal TDHF approach or a semi-classical BNV transport model, incorporating the impact of residual two-body correlations beyond the MF description, in the latter case. A comparative study of semi-classical and quantal approaches within the same EDF framework has enabled us to evaluate how well a semi-classical perspective can capture the properties of nuclear dynamics. Moreover, this approach allowed us to distinguish, for both central and peripheral collisions, the effects of global g.s. deformation, featured in both models, from those arising from  structure details of quantal (shell) nature.

Beyond the well-established role of the isovector channel (and associated symmetry energy), our analysis reveals intriguing connections between the primary features of DD emission and other components of the nuclear effective interaction, including momentum-dependent and surface terms. Notably, incorporating the momentum dependence helps to address the underestimation of centroid energy predicted by theoretical BNV calculations with MI interactions, discussed in Ref.~\cite{parascandoloPRC2022}.
Conversely, we observe that the DD $\gamma$-multiplicity is not much affected by the momentum-dependent terms, nor by the deformation of the $^{152}$Sm g.s. configuration. However, in this work, we highlighted a strong sensitivity of the DD $\gamma$-multiplicity to fine details of the nuclear structure, particularly the nuclear surface of the reactant systems, especially in (mid-)peripheral collisions. Our analysis shows that the magnitude of the power spectrum increases as the radial density profiles of the colliding nuclei (especially the one of the lighter system) become more diffuse, with a smoother transition between bulk and surface region. In other words, enhanced surface effects, driven
by the shape of the g.s. density profiles, favor the DD. In particular, our calculations show that DD $\gamma$-multiplicities closer to the experimental data would be obtained by considering the nuclear density profiles predicted by HF (with respect to the semi-classical calculations). However, it should be noted that the experimental method for deducing the DD prompt $\gamma$-radiation usually involves subtracting the $\gamma$-ray spectra of two different channels, such as the charge-asymmetric reaction considered here and the nearly charge-symmetric reaction $^{48}$Ca + $^{144}$Sm (for which the initial dipole moment almost vanishes)~\cite{parascandoloPRC2022}. This procedure could induce some systematic uncertainties, affecting the comparison with the theoretical predictions of the DD emission.

Finally we note that, since the DD emission acts as a cooling mechanism, enhancing the survival probability of the composite di-nuclear system against fission, our findings  could be relevant for the search of super-heavy element synthesis in low-energy nuclear reactions.

\section*{Acknowledgments}
The authors gratefully acknowledge Guillaume Scamps for his valuable comments and suggestions on this work. This research was supported by the National Natural Science Foundation of China (Grant No. 11905120).

\bibliographystyle{apsrev4-2}
\bibliography{ref}

\newpage

\appendix
\onecolumngrid 

\section{Quadrupole moment of a spheroidal nucleus} 
\label{app:quadrupole}
The quadrupole moment of a spheroidal nucleus, whose surface is described by Eq.~\eqref{spheroideq}, according to Eq.~\eqref{qmeq1}, is \begin{equation}
Q_{20}=\int d^3r (2z^2-x^2-y^2)\rho({\bf r}).
\end{equation}
Assuming a uniform density distribution and taking cylindral coordinates, one then obtain
\begin{eqnarray}
Q_{20}&=&\rho \int_{-a}^{a}dz\int_{0}^{b\sqrt{1-\frac{z^2}{a^2}}}dr r(2z^2-r^2)\int_0^{2\pi} d\varphi \nonumber\\
&=& \frac{8}{15}\rho \pi ab^2\left(a^2-b^2 \right) \nonumber\\
&=& \frac{2}{5}\rho V \left(a^2-b^2\right) \nonumber\\
&=& \frac{2}{5} A r_{0}^2 \left[\left(1+s\right)^2-\left(1-s\right)^2\right] \nonumber\\
&=& \frac{8}{5}Ar_{0}^{2} s
\end{eqnarray}
taking into accont that the volume $V$ of the spheroidal nucleus is
$V = \frac{A}{\rho} = \frac{4}{3}\pi ab^2$.

\section{Multipole moments and deformation parameters}
\label{app:relations}
Let us derive the relations between multipole (mass) moments and deformation parameters
for the shapes that are symmetric with respect to the $\left(y,z\right)$, $\left(x,z\right)$, and $\left(x,y\right)$ planes. This implies that the multipoles $q_{\lambda \mu}$ associated with the corresponding spherical harmonics $Y_{\lambda \mu}\left(\theta,\varphi\right)$ entering the expression for $r\left(\theta,\varphi\right)$ satisfy the following conditions: $q_{\lambda \mu} = q_{\lambda -\mu}$ and $q_{\lambda \mu} = 0$ for any odd $\lambda$ or $\mu$. Specifically, let us consider shapes with axial symmetry about the $z$-axis. In this case, the expression for $r\left(\theta,\varphi\right)$ has following form
\begin{equation}
r\left(\theta ,\varphi\right) = C r_{0}\left(1+\beta_2Y_{20}+\beta_4Y_{40}+\beta_6Y_{60}\right)
\end{equation}
where $C$ is given by
\begin{eqnarray}
C^{-3} & =& 1+\frac{3}{4\pi} (\beta _2^2+\beta _4^2+\beta _6^2) +\frac{\sqrt 5}{28\pi^{3/2}} \beta _{2}^{3}+\frac{9}{28\pi^{3/2}} \beta _{2}^2\beta_{4} +\frac{15\sqrt 5}{154\pi^{3/2}} \beta_{2}\beta_{4}^{2} +\frac{243}{4004\pi ^{3/2}} \beta_{4}^{3}  +\frac{45\sqrt{5}}{44\sqrt{13}\pi ^{3/2}} \beta_{2}\beta_{4} \beta_{6} \nonumber \\
&&+\frac{15}{22\sqrt{13}\pi ^{3/2}} \beta_{4}^{2} \beta_{6} +\frac{21}{44\sqrt{5}\pi^{3/2}} \beta_{2} \beta_{6}^{2} +\frac{63}{374\pi ^{3/2}} \beta _{4} \beta_{6}^{2}+\frac{50\sqrt{13}}{3553\pi^{3/2}}\beta_{6}^{3}
\end{eqnarray}
as determined from the volume conservation condition.

Moreover, the multipole (mass) moment operator $\hat{Q}_{\lambda 0}$ is defined as~\cite{Cwiok:1996kn}
\begin{equation}
\hat{Q}_{\lambda 0}=2r^{\lambda } P_{\lambda }\left(\cos{\theta}\right)=2r^{\lambda }\sqrt{\frac{4\pi}{2\lambda+1}} Y_{\lambda 0}\left(\theta ,\varphi \right)
\end{equation}
where $P_{\lambda }$ are the standard Legendre polynomials. To determine the corresponding expectation values, the integration over the spatial coordinates must be performed. Specifically, the integration over $r$ and $\varphi$ yields the following expression:
\begin{equation}
Q_{\lambda 0}=\frac{4\pi }{\lambda + 3}C^{\lambda + 3} r_{0}^{\lambda + 3} \int_0^{\pi }\left(1+\beta_2Y_{20} +\beta_{4} Y_{40}  + \beta_{6} Y_{60} \right)^{\lambda + 3} \sqrt{\frac{4\pi}{2\lambda+1}}Y_{\lambda 0}\left(\theta\right) \sin{\theta}\mathit{d\theta}
\end{equation}
Finally, after integrating over $\theta$, we expand the resulting expressions into a power series of $\beta_{i}, {i = 2, 4, 6}$. By retaining only terms up to the second order, we finally get the following expressions:
\begin{eqnarray}
Q_{20} & = &\frac{3}{\sqrt{5\pi }} Ar_0^2 \left(\beta _2+\frac 2 7\sqrt{\frac 5{\pi }}\beta _2^2+\frac{12}{7\sqrt{\pi }} \beta _2\beta_4 +\frac{20}{77}\sqrt{\frac 5{\pi }}\beta_{4}^{2}+\frac{30}{11}\sqrt{\frac 5{13\pi }}\beta _4\beta_6+\frac{14}{11\sqrt{5\pi }}\beta _6^2\right) \label{eq:q20_app}\\
Q_{40} & = & \frac{1}{\sqrt{\pi }} Ar_{0}^{4} \left(\beta _4+\frac{9}{7\sqrt{\pi}} \beta_{2}^{2} +\frac{60}{77}\sqrt{\frac 5{\pi}}\beta_{2} \beta_{4}+\frac{729}{1001\sqrt{\pi}} \beta_{4}^{2} +\frac{45}{11}\sqrt{\frac 5{13\pi}}\beta _2\beta_6+\frac{60}{11\sqrt{13\pi}} \beta_{4}\beta_{6} +\frac{126}{187\sqrt{\pi}} \beta_{6}^{2}\right)\\
Q_{60} & = & \frac{3}{\sqrt{13\pi }} Ar_{0}^{6} \left(\beta_{6} + \frac{60}{11}\sqrt{\frac 5{13\pi}} \beta_{2}\beta_{4} + \frac{40}{11\sqrt{13\pi}} \beta_{4}^{2} +\frac{56}{11\sqrt{5\pi}}\beta_{2}\beta_{6}  + \frac{336}{187\sqrt{\pi}} \beta_{4} \beta_{6} + \frac{800\sqrt{13}}{3553\sqrt{\pi}}\beta_{6}^{2}\right)
\end{eqnarray}
Equation~\eqref{eq:q20_app} is thus equivalent to Eq.~\eqref{qmeq2}, except for additional terms that depend on $\beta_{6}$ and correspond to $Y_{60}$ deformations, which were not considered in Ref.~\cite{Cwiok:1996kn}.

\end{document}